\documentclass[openacc]{rstransa}
\usepackage{siunitx} 
\usepackage{color}
\usepackage[normalem]{ulem}

\definecolor{darkgreen}{rgb}{0,0.6,0}

\begin{document}

\title{Recent developments of turbulent emulsions in Taylor--Couette flow}

\author{
Lei Yi$^{1}$, Cheng Wang$^{1}$, Sander G. Huisman$^{2}$, and Chao Sun$^{1,2,3}$}

\address{$^{1}$Center for Combustion Energy, Key Laboratory for Thermal Science and Power Engineering of Ministry of Education, Department of Energy and Power Engineering, Tsinghua University, 100084 Beijing, China\\
$^{2}$Physics of Fluids Group, Max Planck UT Center for Complex Fluid Dynamics, University of Twente, 7500 AE Enschede, Netherlands\\
$^{3}$Department of Engineering Mechanics, School of Aerospace Engineering, Tsinghua University, Beijing 100084, China}

\subject{fluid mechanics}


\corres{Chao Sun\\
\email{chaosun@tsinghua.edu.cn}}

\begin{abstract}
Emulsions are common in many natural and industrial settings. Recently, much attention has been put on understanding the dynamics of turbulent emulsions. This paper reviews some recent studies of emulsions in turbulent Taylor--Couette flow, mainly focusing on the statistics of the dispersed phase and the global momentum transport of the system. 
We first study the size distribution and the breakup mechanism of the dispersed droplets for turbulent emulsions with a low volume-fraction (dilute) of the dispersed phase. 
For systems with a high volume-fraction of the dispersed phase (dense), we address the detailed response of the global transport (effective viscosity) of the turbulent emulsion and its connection to the droplet statistics. 
Finally, we will discuss catastrophic phase inversions, which can happen when the volume fraction of the dispersed phase exceeds a critical value during dynamic emulsification. We end the manuscript with a summary and an outlook including some open questions for future research. This article is part of the theme issue `Taylor--Couette and Related Flows on the Centennial of Taylor's Seminal Philosophical Transactions Paper'.
\end{abstract}


\begin{fmtext}
\end{fmtext}
\maketitle

\section{Introduction}
Flow confined between two coaxial, independently-rotating cylinders was first studied by Maurice Couette and Geoffrey I.\ Taylor and the flow geometry now bears their name: Taylor--Couette (TC).
Maurice Couette described this kind of arrangement to study the rheology of fluids in his thesis in 1890~\cite{couette1890etudes}.
In 1923, Taylor gave the mathematical description of the stability of TC flow~\cite{taylor1923viii}, which was confirmed by both the remarkable experiments by Taylor himself and those done by Lewis~\cite{lewis1928experimental}.
TC flow is one of the paradigmatic systems in the field of fluid mechanics due to its simple, closed geometry with well-defined boundary conditions, resulting in accessible experiments and numerical simulations, and making the flow theoretically tractible.
Over the past century, this system has attracted great attention in various fields of fluid mechanics, including instabilities~\cite{synge1938stability,coles1965transition,busse1967stability}, flow patterns~\cite{andereck1986flow,cross1993pattern,fardin2014hydrogen}, and turbulence~\cite{siggia1994high,lathrop1992turbulent,grossmann2016high}, among others.
In many natural and industrial processes, turbulent liquid flows contain dispersed droplets, bubbles, or particles~\cite{toschi2009lagrangian,voth2017anisotropic,mathai2020bubbly}.
Turbulent multiphase flows in TC systems are investigated in various studies, including bubbly flows~\cite{van2005drag,van2007bubbly,verschoof2016bubble}, particle-laden flows~\cite{bakhuis2018finite,dash2020particle,singh2022counter,wang2022finite}, and liquid-liquid dispersed flows~\cite{lemenand2017turbulent}.
The flow of emulsions (i.e.\ liquid-liquid dispersed flow) is common in nature and various industrial settings, such as oil recovery, chemical engineering, pharmaceuticals, and food processing~\cite{kilpatrick2012water,wang2007oil,mcclements2007critical}.
Although the applications of the emulsions are wide, the investigation of the underlying physics of emulsions, particularly turbulent emulsions, is still limited.

For very low volume-fractions, turbulent emulsions can be characterized by the break-up of the dispersed droplets, as the chance of coalescence is negligible. In this case, the droplet size is determined by the turbulent flow, while the effect of the dispersed droplets on the macroscopic properties is relatively minor. 
For emulsions with a high volume-fraction of the dispersed phase, the interaction of the microscopic droplets and the macroscopic flow dynamics governs the statistical properties of the turbulent emulsion. 

Due to its importance, various numerical studies have been carried out to study emulsions in a variety of flow configurations, e.g.\ homogeneous and isotropic turbulence \cite{mukherjee2019droplet, crialesi2022modulation} and planar Couette flow \cite{rosti2018rheology, de2019effect, de2020numerical, rosti2021shear}. However, to the best of the authors' knowledge, few have reported results based on experimental measurements, especially in highly turbulent conditions where performing accurate measurements is of tremendous difficulty. 
Going through the full topic of turbulent emulsions in this limited review is not feasible, thus our attention is mainly limited to the recent development in experimental studies of turbulent TC emulsions.
We use some of our work as examples to explain the developments and challenges of this newly-emerged field.
Some important relevant experimental and numerical studies on turbulent emulsions in related flows will also be included for comparison.

In this review, we aim to discuss the dynamics of emulsions in turbulent TC flow, mainly focusing on two aspects: the microscopic droplet formation and the macroscopic properties of the global transport (effective viscosity).
For that purpose, we first show the size distribution of dispersed droplets and discuss the droplet break-up mechanism for low volume-fractions of the dispersed phase.
We then try to understand how a high volume-fraction of the dispersed phase affects the global transport of the emulsion using the concept of an effective viscosity, which includes a discussion of shear thinning effects.
Next, we show experimental results of the catastrophic phase inversion of the turbulent emulsion in TC flow and other flow systems, focusing on the associated dramatic change of rheological properties of the mixture.
The review ends with a summary and an outlook.

\section{Typical TC setup for turbulent emulsion studies and relevant parameters}

\begin{figure}
    \centering
    \includegraphics[width = 1\textwidth]{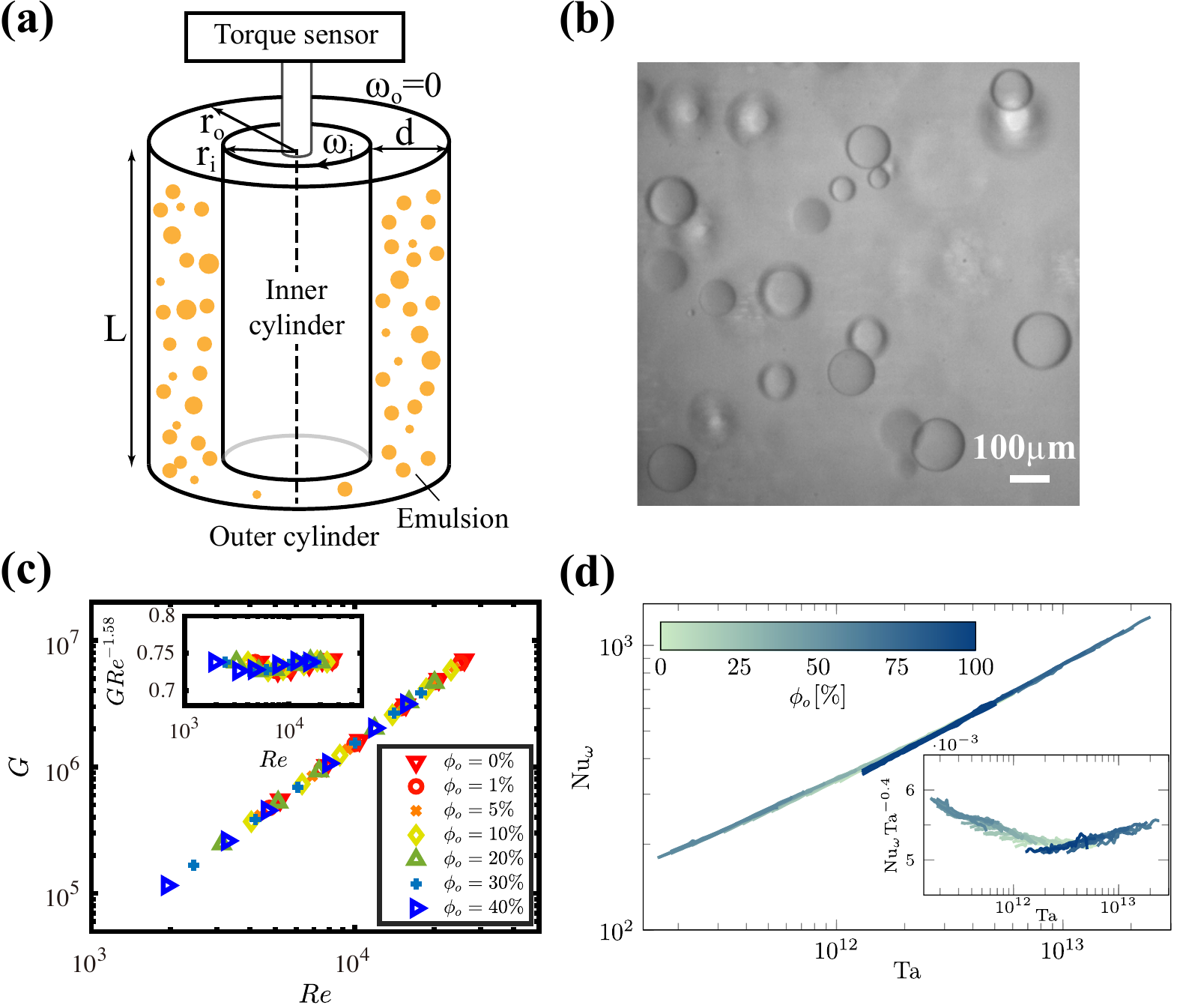}
    \caption{
        (a) Typical sketch of an experimental TC set-up. By rotating the inner cylinder at a constant angular velocity $\omega_{i}$, while maintaining the outer cylinder fixed, an emulsion is formed in the gap between the cylinders. The torque on the inner cylinder is measured by a torque sensor. 
        (b) A typical snapshot of the oil droplet dispersion in a turbulent TC flow. Here, the oil volume fraction is $\phi_{o} = 1\%$, and $\text{Re} = 10^{4}$.
        (c, d) The global transport of the system ($G$ or Nu) can be expressed to be a power-law dependence with the driving parameter (Re or Ta) with a local effective scaling exponent:
        (c) The measured $G$ versus $\text{Re}$ for various volume-fractions in the Tsinghua TC system. The inset shows the dimensionless torque compensated with an effective scaling exponent; 
        (d) The measured $\text{Nu}_{\omega}$ versus $\text{Ta}$ for various volume-fractions in the Twente TC system~\cite{gils2011b}. The inset shows the compensated datasets with an effective scaling exponent. 
        Figures (a--c) adapted from Yi {\it et al.}~\cite{yi2021global}; Figure (d) adapted from Bakhuis {\it et al.}~\cite{bakhuis2021catastrophic}
    }
    \label{fig1} 
\end{figure}

The driving of the TC system is quantified by the Reynolds number: 
\begin{align}
    \text{Re} &= \frac{\omega_{i}r_id}{\nu_{w}},
\end{align}
where $r_i$ ($r_o$) is the radius of the inner (outer) cylinder, $d=r_o-r_i$ the size of the gap, $\omega_i$ the rotational velocity of the inner cylinder, and $\nu_w$ the viscosity of water.
The global response of the system can be characterized by the dimensionless torque:
\begin{align}
    G &= \frac{T}{2\pi L\rho_{w}\nu_{w}^{2}},
\end{align}
 where $T$ is the torque needed to sustain a constant angular velocity of the inner cylinder.
 Instead of $\text{Re}$ and $G$, alternatively one can use another pair of parameters---inspired by the analogy of TC and Rayleigh--B\'enard convection---to describe the driving and the response of the system, they are the Taylor and Nusselt numbers:
\begin{align}
\text{Ta} &= \frac{(1+\eta)^{4}}{64 \eta^{2}} \frac{\left(r_{\mathrm{o}}-r_{\mathrm{i}}\right)^{2}\left(r_{\mathrm{i}}+r_{\mathrm{o}}\right)^{2}\left(\omega_{\mathrm{i}}-\omega_{\mathrm{o}}\right)^{2}}{\nu^{2}}, 
\end{align}
\begin{align}
\text{Nu}_{\omega} &= J^{\omega} / J_{\text {lam }}^{\omega},
\end{align}
for which the latter is the ratio of the angular velocity flux $J^{\omega}$ and its value $J_{\mathrm{lam}}^{\omega}=2 \nu r_{\mathrm{i}}^{2} r_{\mathrm{o}}^{2}\left(\omega_{\mathrm{i}}-\omega_{\mathrm{o}}\right) /\left(r_{\mathrm{o}}^{2}-r_{\mathrm{i}}^{2}\right)$ for the laminar, non-vortical case.

Emulsions consists of two immiscible liquids---most often oil and water---of which the composition can be determined by the volume-fraction of oil: 
\begin{equation}
    \phi_{o} = \frac{V_{o}}{V_{o} + V_{w}},
\end{equation}
where $V_{o}$ and $V_{w}$ denote the volume of the oil phase and aqueous phase in the TC system, respectively.

In this paper, we will mainly discuss the results of two TC setups, i.e.\ the setup at Tsinghua University used in Yi {\it et al.}~\cite{yi2021global,yi2022physical} and the setup at the University of Twente used in Bakhuis {\it et al.}~\cite{bakhuis2021catastrophic}.
Figure~\ref{fig1}(a) shows a sketch for studying the emulsions in a turbulent TC flow.
The Tsinghua-TC system has an inner cylinder with an outer radius $r_{i}=25\rm~mm$, and a transparent outer cylinder with inner radius $r_{o}=35\rm~mm$, giving a gap $d=10\rm~mm$ and a radius ratio $\eta=r_{i}/r_{o}=0.71$. The height of the cylinder is $L=75\rm~mm$ resulting in an aspect ratio $\Gamma=L/d=7.5$.
The outer cylinder and end plates are stationary while the inner cylinder rotates at a constant angular velocity $\omega_{i}$. Thus, a turbulent TC flow is formed and an emulsion is formed in the gap. During the experiments, the working temperature was maintained at $(22\pm0.1)^\circ$C using a circulating water bath system. The Reynolds number range of the Tsinghua-TC is from $10^{3}$ to $3\times 10^{4}$. In the experiments by Yi {\it et al.}~\cite{yi2021global,yi2022physical}, an aqueous mixture of ultra-pure water and ethanol ($\rho_{w}=860\rm~kg/m^{3}$, $\nu_{w}=2.4\times10^{-6}\rm~m^{2}/s$) and silicone oil (density $\rho_{o}=866\rm~kg/m^{3}$, and viscosity $\nu_{o}=2.1 \times10^{-6}\rm~m^{2}/s$) are used. 
The Twente $\text{T}^3\text{C}$ (Twente Turbulent Taylor--Couette) facility has an inner cylinder radius $r_i=\SI{200}{\milli\meter}$, an outer cylinder radius $r_o=\SI{279.4}{\milli\meter}$, and a height $L=\SI{927}{\milli\meter}$, resulting in a gap width $d=\SI{79.4}{\milli\meter}$, a radius ratio $\eta=0.716$, and an aspect ratio $\Gamma = 11.7$. The temperature of the fluids is kept within $(21\pm0.5)^\circ$C. The Taylor number range of $\text{T}^3\text{C}$ is from $10^{11}$ to $3\times 10^{13}$, which corresponds to Reynolds number in the range of 10$^5$ to 2$\times$10$^6$. In the experiments by Bakhuis {\it et al.}~\cite{bakhuis2021catastrophic}, the demineralized water and a low-viscosity silicone oil with $\nu_o=\SI{1.03}{\milli\meter\squared\per\second}$ are used. 
In both systems torque sensors are used to measure the torque exerted onto the inner cylinder with high precision.
The measured dimensionless torque data as a function of Re and Ta are shown in Figure~\ref{fig1}(c,d), which will be discussed below in Section 3. 
High-speed cameras were used to analyze the dispersed droplets in the emulsion.

\section{Results}

\subsection{The dispersed phase statistics in dilute turbulent emulsions}

\begin{figure}
    \centering
    \includegraphics[width = 1\textwidth]{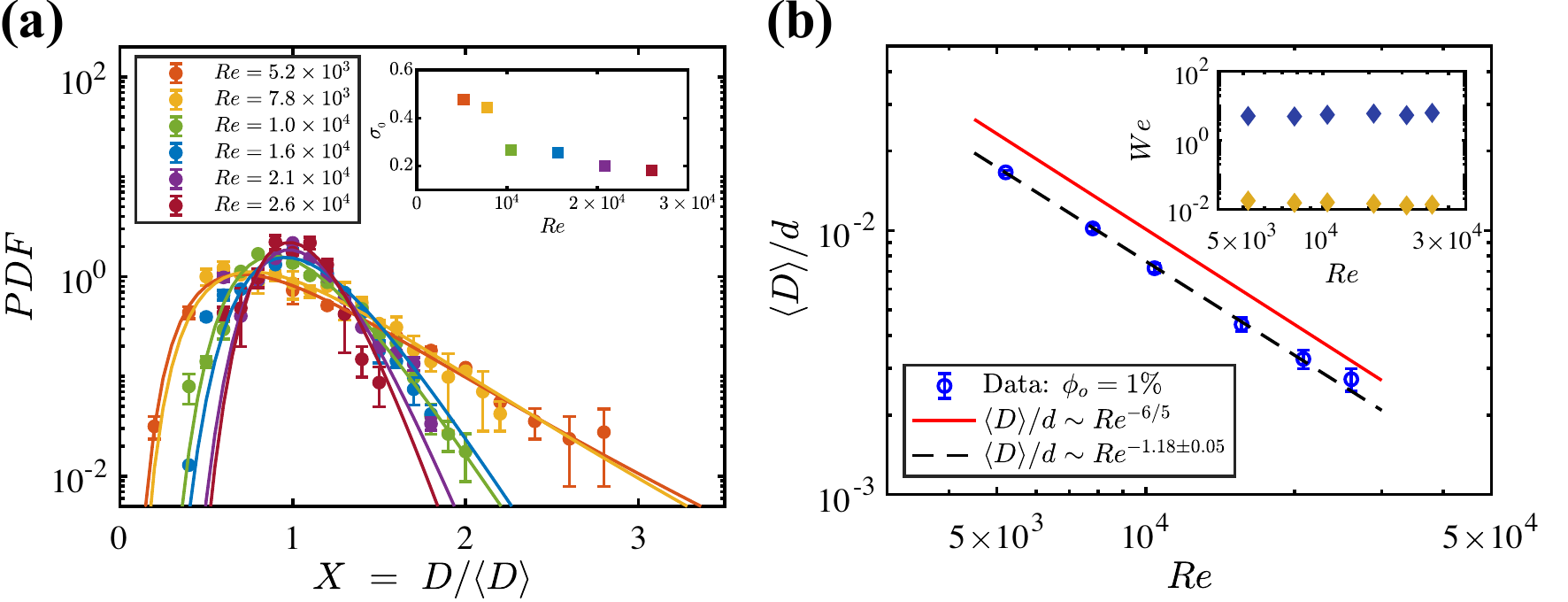}
    \caption{
    (a) The probability density function (PDF) of the droplet size. The solid lines show the fitting results with a log-normal distribution function (Eq. \ref{log-normal}). The statistical error bars are calculated based on $\mathcal{O}(10^{3})$ droplet samples for each $Re$ in the experiments. Inset: the fitted values of the standard deviation $\sigma_{0}$ versus $Re$. 
    (b) The average droplet diameter normalized by the gap width as a function of the Reynolds number.
	The blue open circles are the experimental data at the oil volume-fraction of $1\%$, and the error bars are based on the image analysis.
	The black dashed line denotes the weighted fit of the experimental data, and the red solid line shows the $-6/5$ scaling dependence. Inset: The calculated Weber number as a function of Re for the two theoretical approaches. The yellow diamonds are the results from K-H approach, while the blue diamonds are from the Levich's approach.
	Figure (a,b) adapted from Yi {\it et al.}~\cite{yi2021global}.
}.
    \label{fig2} 
\end{figure}

The size of the dispersed droplets in an emulsion is a very important parameter, which can be connected to the stability and the rheology of the emulsion, and determines the interfacial area for mass transfer.

Here, we present some experimental results of the droplet statistics in a low volume-fraction ($\phi_{o} = 1\%$) studied in the work by Yi {\it et al.} \cite{yi2021global}.
As shown in figure~\ref{fig1}(b), the snapshots of the dispersed oil droplets in the flow are captured using a high-speed camera, and the droplet size has a wide distribution. 
We first focus on the droplet size distribution at various Reynolds numbers. The droplets interface can be extracted from the recorded images using image analysis, and then the diameters of the detected droplets can be obtained.
The droplet diameter is normalized with the average droplet diameter as $X
= D/\left<D\right>$. The probability density function (PDF) of the droplet size is computed as a function of $X$ for various Reynolds numbers (see figure~\ref{fig2}(a)). It is found that the droplet size distribution can be well described by a log-normal distribution:
\begin{align}
    P(X) &= \frac{1}{X\sigma_0\sqrt{2\pi}}\exp\left[-\frac{\left[ \log(X) - \log(X_0) \right] ^2}{2\sigma_0^2}\right],
    \label{log-normal}
\end{align}
where $X_0$ and $\sigma_0$ are fitting parameters corresponding to the mean and the standard deviation of the distribution after taking the log of the variable, respectively.
These log-normal distributions indicate that the droplet generation in the present system could be dominated by the cascade of the random fragmentation process.
The investigation of this process can be traced back to Kolmogorov, who gave a breakup theory for solid particles that describes the cascade of uncorrelated breakage events, leading to the log-normal distribution of the particle size~\cite{kolmogorov1941log}.
The log-normal distribution has been widely used to describe the fragment size in the spray community~\cite{gorokhovski2003analyses,villermaux2007fragmentation,xu2022droplet}.
However, it is remarkable to observe the validity of this theory in the breakup of liquid droplets in a turbulent TC flow.
The width of the distribution of the droplet size is found to decrease monotonically with increasing $\text{Re}$, which is confirmed by the decreasing trend of the fitted value of the standard deviation $\sigma_0$ (see the inset in figure~\ref{fig2}(a)), suggesting that the drops become more uniformly distributed with increasing turbulence level of the system.
The next question we focus on is what sets the droplet size in the current system.

Droplet breakup in a turbulent flow is widely investigated using the Kolmogorov--Hinze (KH) theory, in which the droplet breakup is governed by the competition between the resisting interfacial tension and the deforming external dynamic pressure force (turbulent fluctuations) over the droplet size~\cite{kolmogorov1949breakage,hinze1955fundamentals}. 
The ratio of these two forces is usually indicated by the droplet Weber number $\text{We} = \rho \overline{\delta u^2} D/\gamma$, 
where $\rho$ is the density of the continuous phase, $\overline{\delta u^2}$ the mean-square velocity difference over the distance of one droplet diameter $D$, and $\gamma$ the interfacial tension between the two liquids.
As the droplet size falls in the inertial turbulent sub-range, this gives $\overline{\delta u^2} = 2(\varepsilon D)^{2/3}$ ($\varepsilon$ the local energy dissipation rate) according to Batchelor's theory~\cite{batchelor1953theory}, yielding a Weber number $\text{We} = 2\rho\varepsilon^{2/3}D^{5/3}/\gamma$.
Various studies have shown the existence of an order-of-unity critical value of the Weber number (i.e. $\text{We}\sim\mathcal{O}(1)$) over which breakup occurs~\cite{hesketh1991bubble,risso1998oscillations,lemenand2017turbulent}.
The maximum stable droplet size in a homogeneous and isotropic turbulent flow is given by $D_{\text{max}} = C(\gamma/\rho)^{3/5}/\varepsilon^{2/5}$,
where $C$ is a constant coefficient, which is the main result of the work by Hinze~\cite{hinze1955fundamentals}.
Moreover, it has been found that the average droplet diameter, $\left<D\right>$, can be used as the characteristic size in the KH prediction~\cite{boxall2012droplet,perlekar2012droplet,lemenand2003droplets}.
After that, the KH theory has been used in various experimental and numerical studies~\cite{risso1998oscillations,perlekar2012droplet,eskin2017modeling,rosti2019droplets}.
However, it has already been pointed out that the KH theory has some limitations, especially in non-homogeneous turbulent flows~\cite{hinze1955fundamentals,lemenand2017turbulent}.

First, one considers that the droplet size could be determined by the turbulent fluctuations in the bulk flow of the system, where most droplets distribute. The local energy dissipation rate in the bulk area of a TC flow can be estimated by $\varepsilon_b \sim u_{_T}^{3}/\ell$, where $u_{_T}$ is the typical velocity fluctuation and $\ell$ the characteristic length scale of the flow~\cite{ezeta2018turbulence}. By using $u_{_T} \sim \omega_{i}r_{i} \sim \text{Re} \nu /d$ and $\ell \sim d$~\cite{van2012optimal}, $\varepsilon_b \sim \text{Re}^3\nu^3/d^4$ is obtained and thus a scaling dependence,
\begin{align}
    \frac{\left<D\right>}d &\propto \text{Re}^{-6/5},
\end{align}
based on the KH prediction. This agrees well with the results in experiments (see figure~\ref{fig2}(b)).
However, a further quantitative study should consider the Weber number (comparing inertia and surface tension forces), which is expressed as 
\begin{align}
    \text{We} = \frac{2\rho_{w}\varepsilon_b^{2/3}D^{5/3}}\gamma.
\end{align}
Here one can estimate the bulk dissipation rate as $\varepsilon_b \approx 0.1T\omega_{i}/[\pi(r_{o}^{2}-r_{i}^{2})L\rho_{w}]$ in the present system~\cite{ezeta2018turbulence}.
As shown by the yellow diamonds in the inset of figure~\ref{fig2}(b), the result of the Weber number ranges between 0.013 to 0.018, two orders of magnitude smaller than the critical value (i.e. $\text{We}\sim\mathcal{O}(1)$), suggesting that the deforming external force due to the turbulent fluctuations is much smaller than the resisting interfacial tension. This indicates that the droplet size is not determined by the turbulent fluctuations in the bulk of the system, where most droplets exist.

Indeed, it has already been found that the droplet breakup usually occurs at the place where the most intense stress contributes to the deformation~\cite{hesketh1991experimental,afshar2013liquid}. Thus, the droplet size is expected to be determined by the turbulence inside the boundary layer, in which the KH theory has some limitations. The breakup of the droplet in a non-homogeneous turbulent flow near the wall was investigated by Levich~\cite{levich1962physicochemical}, who considered the deformation stress as the dynamic pressure difference over the droplet size close to the wall using a logarithmic distribution of the mean velocity in a turbulent boundary layer. Based on Levich's theory, the droplet diameter can be given as 
\begin{align}
    \left<D\right> &= 2\sqrt{\frac{\gamma \nu_{w}}{25\rho_{w} u_{\ast}^{3}}}, 
\end{align}
where $u_{\ast} = \sqrt{\tau_w/\rho_{w}} = \sqrt{T/(2\pi\rho_{w} r_{i}^{2}L)}$ is the shear velocity. Then, the droplet diameter is found to have a scaling dependence on the Reynolds number as
\begin{align}
\frac{\left<D\right>}d &\propto \text{Re}^{-1.19},
\end{align}
which is very close to the $-6/5$ scaling prediction of the KH theory and agrees with experimental results. Here, Yi {\it et al.} \cite{yi2022physical} used the effective scaling of $G \propto \text{Re}^{1.58}$ obtained in their experiments.
The Weber number here can be calculated as the ratio of the dynamic pressure difference caused by the gradient of the mean flow to the interfacial tension: 
\begin{align}
\text{We} = \frac{25\rho_{w} u_{\ast}^{3}\left<D\right>^{2}}{2\nu_{w}\gamma},
\end{align}
of which the result is around 5 (see the blue diamonds in the inset of figure~\ref{fig2}(b)). 
This value is consistent with the critical value of the breakup of the droplet in a turbulent flow, suggesting a comparable balance between the deforming force and resisting interfacial tension over the droplet size~\cite{risso1998oscillations,lemenand2017turbulent}.
Moreover, the boundary layer thickness is found to be at least 5 times the average droplet diameter, which also validates the above-mentioned discussion.
Thus, it is concluded that the droplet size in the current system is determined by the dynamic pressure difference induced by the gradient of mean velocity in the boundary layer as proposed by Levich \cite{levich1962physicochemical}.

\subsection{Global transport statistics in dense turbulent emulsion systems}
For high volume-fractions, turbulent emulsions becomes more complex as the coupling interaction between the local dispersed droplets and the global flow (rheology) dominates the dynamics. 
Previous studies have shown that the viscosity of the fluid can change dramatically due to the addition of solid particle, droplets, or bubbles~\cite{van2005drag,stickel2005fluid,guazzelli2018rheology}.
It has been found that the viscosity of an emulsion increases with increasing dispersed phase volume-fraction~\cite{pal1992rheology,rosti2021shear}.
However, most investigations of the viscosity of the dispersed suspension is usually performed in the laminar flow regime using conventional rheometers.
Determining the rheological properties of an emulsion in a turbulent flow is challenging and related studies are still limited, especially for high volume-fractions~\cite{pal1992rheology,faroughi2015generalized,de2019effect}.

\begin{figure}
    \centering
    \includegraphics[width = 1\textwidth]{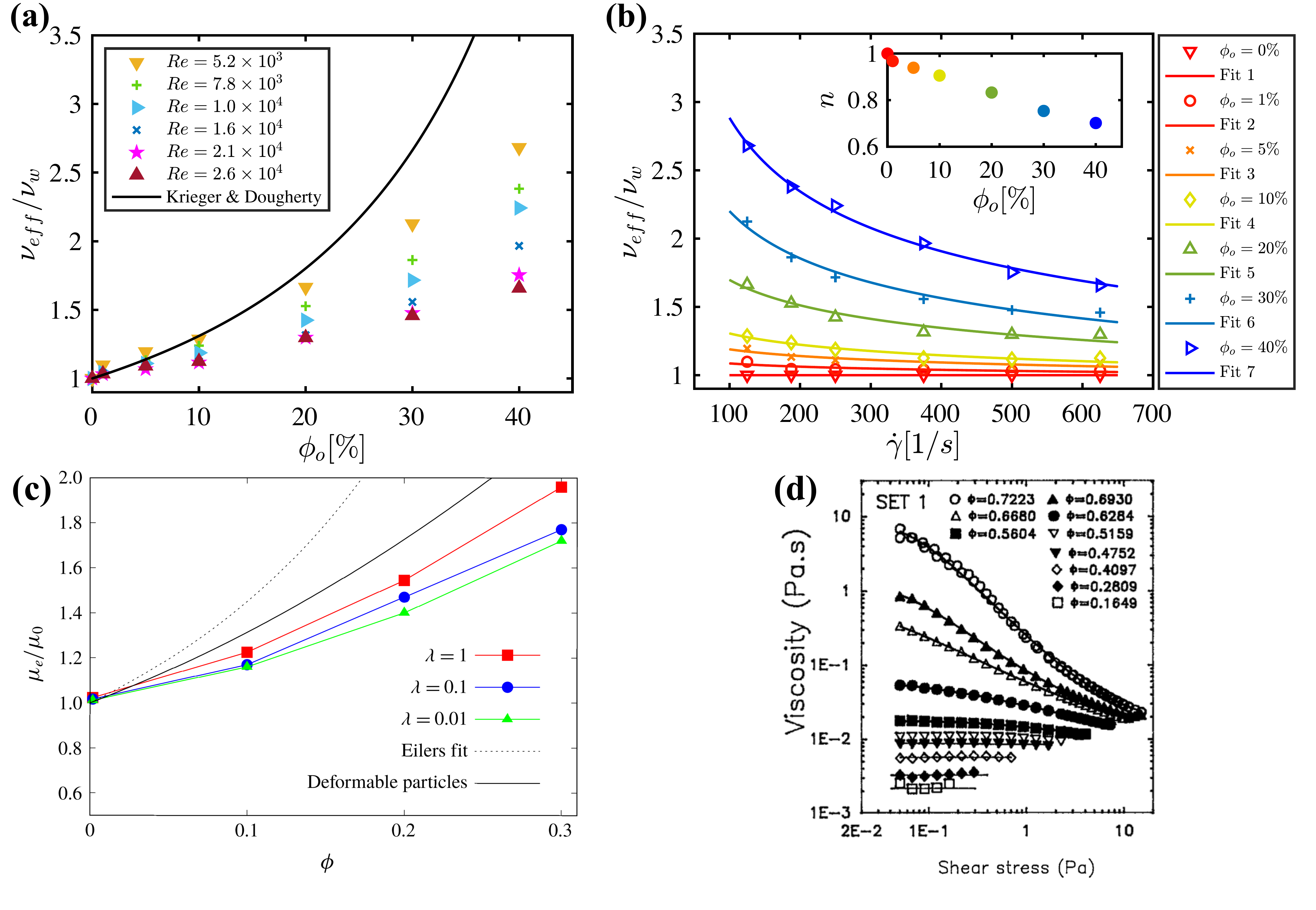}
    \caption{
    (a) The normalized effective viscosity $\nu_{\text{eff}}/\nu_{w}$ as a function of the oil volume-fraction $\phi_{o}$ at various $\text{Re}$. The solid black line represents the result of the viscosity model for solid particle suspensions~\cite{krieger1959mechanism}. (b) The normalized effective viscosity versus the characteristic shear rate $\dot{\gamma}$ of the flow. Fitting results using the Herschel--Bulkley model \cite{herschel1926konsistenzmessungen} are denoted by solid lines for various oil volume-fractions. The inset gives the values of the flow index $n$ as the fitting results. Figures (a,b) adapted from Yi {\it et al.}~\cite{yi2021global}; (c) The effective viscosity of an emulsion in a planar Couette flow obtained at various viscosity ratios and volume fractions when the droplet coalescence is prohibited. $\lambda$ is the viscosity ratio of dispersed droplet to the continuous phase. The line denoted as `deformable particles' corresponds to the fit proposed in Rosti $\&$ Brandt ~\cite{rosti2018rheology} for initially spherical viscoelastic particles. The line denoted as `Eilers fit' is given by Eilers~\cite{eilers1941viskositat}. Figure adapted from De Vita {\it et al.}~\cite{de2019effect}, (d) The measured viscosity of an emulsion with two immiscible liquids versus shear stress. Figure adapted from Rajinder Pal~\cite{Pal2000ShearVB}.
    Note that the $\phi$ in panels (c,d) is the same as the oil volume-fraction $\phi_{o}$ defined in this review.
    }
    \label{fig3} 
\end{figure}

\textcolor{black}{A method was recently developed by Bakhuis {\it et al.}~\cite{bakhuis2021catastrophic} to measure the effective viscosity of a turbulent emulsion in a TC flow. 
As shown in figure~\ref{fig1}(d), they assumed that the scaling dependence of $\text{Nu}_{\omega} \propto \text{Ta}^{\alpha}$ (or $G \propto \text{Re}^{\beta}$) found in a single-phase TC flow can still be applied to a turbulent emulsion, which was confirmed by experimental results (see figure~\ref{fig1}(c,d))~\cite{grossmann2016high,bakhuis2021catastrophic}. This scaling can then be exploited to determine the effective viscosity of an emulsion (see figure~\ref{fig4}b), which characterizes the global transport of the system.
Using the same approach as proposed by Bakhuis {\it et al.}~\cite{bakhuis2021catastrophic}, Yi {\it et al.}~\cite{yi2021global} calculated the effective viscosity of an emulsion at various $\phi_{o}$ and $\text{Re}$ based on the measured scaling dependence of $G \propto \text{Re}^{\beta}$ as shown in figure~\ref{fig1}(c). 
The calculated results of the effective viscosity are shown in figure~\ref{fig3}(a).
The effective viscosity is found to increase with increasing oil volume-fraction, for all $\text{Re}$ explored.}
The stronger dependence of viscosity on the volume-fraction is displayed for larger $\phi_{o}$. This can be attributed to the increasing hydrodynamic interactions or direct contact between droplets in the flow for higher oil volume fractions, which is similar to the dynamics of solid spherical particle suspensions~\cite{guazzelli2018rheology}. 
\textcolor{black}{A similar volume-fraction dependence of the effective viscosity of emulsions has also been reported recently by De Vita {\it et al.}~\cite{de2019effect} (figure~\ref{fig3}(c)). With numerically exerting an Eulerian force to prohibit the droplet coalescence, they found that the effective viscosity of an emulsion in a planar Couette flow is always greater than 1 for various viscosity ratios, suggesting that emulsions behave as suspensions of deformable particles when coalescence is absent.} 

Let's compare the turbulent emulsion results to the typical model for solid particle suspensions (see black line in figure~\ref{fig3}(a))~\cite{krieger1959mechanism}. The values of the model of solid particle suspension are found to be larger than the corresponding results for an emulsion, which could be partially due to the existence of the friction contact between solid surface in the solid particle suspension. In addition, particle (droplet) size distribution has been found to affect the rheology of the particle-suspension (emulsion) system~\cite{ramirez2002drop,shewan2015analytically}. The model used for particle suspensions is based on the assumption of mono-disperse droplet size while the current emulsions have a wide droplet size distribution. One should also be noted that tremendous difference exist between droplets and solid particles. For instance, the shape and size of the solid particles are fixed, but the dispersed oil droplets experience deformation, breakup, and coalescence. Compared to the solid particle suspensions, the droplets dispersed in emulsions are expected to show richer dynamics.

In addition, the effective viscosity is found to decrease with increasing $\text{Re}$, for fixed $\phi_{o}$, suggesting a shear thinning effect.
Here, by taking the planar shear flow at low Reynolds numbers as reference, one can use a characteristic shear rate defined as 
\begin{align}
    \dot{\gamma} &= \frac{\omega_{i}r_i}d,
\end{align}
in the current TC flow. As shown in figure~\ref{fig3}(b), the shear thinning effect is characterized by decreasing viscosity values with increasing the shear rate, which is more pronounced for high volume fractions.
A quantitative description of the shear thinning effect can be given using the Herschel-Bulkley model~\cite{herschel1926konsistenzmessungen}:
\begin{align}
    \mu_{\text{eff}} &= k_{0}|\dot{\gamma}|^{n-1}+\tau_{0}|\dot{\gamma}|^{-1},
\end{align} 
where $\mu_{\text{eff}}$ is the effective dynamic viscosity, $k_{0}$ the consistency, $n$ the flow index, and $\tau_{0}$ the yield shear stress.
Considering that the system is far from a jamming state, the yield shear stress is zero ($\tau_{0}=0$), giving $\nu_{\text{eff}}/\nu_{w}=K|\dot{\gamma}|^{n-1}$, where $K$ is a constant.
The experimental results are found to agree well with the model, in which the decreasing flow index indicates the more pronounced shear thinning effect for high volume-fractions (see the inset in figure~\ref{fig3}(b)).
This agreement opens a possible avenue for the description of the rheological properties of turbulent emulsion using classical non-Newtonian models. In addition, the observed shear thinning effect has potential applications in the drag reduction of the multi-component liquid system in turbulent states. 
This shear thinning effect has also been reported both in the suspensions of deformable particles under a steady shear flow~\cite{adams2004influence,rosti2018rheology,de2019effect} and in emulsions \cite{Pal2000ShearVB} (figure~\ref{fig3}(d)). 
For example, Direct Numerical Simulations (DNSs) have shown that suspensions of deformable particles in planar Couette flow show shear thinning effects due to the particle deformability, where the effective viscosity of the suspension can be described by the well-known Eilers fit with a reduced effective volume-fraction~\cite{rosti2018rheology}.

As shown in figure~\ref{fig3}(d), Rajinder Pal~\cite{Pal2000ShearVB} found that emulsions are Newtonian (or, weakly shear-thinning) at low-to-moderate volume-fractions, while at higher volume-fractions, emulsions appear to be strongly shear-thinning. Additionally, Rajinder Pal~\cite{Pal2000ShearVB} pointed out that the viscosity of shear-thinning emulsions is strongly influenced by the droplet size. 
He found that a significant increase in the viscosity occurs when the droplet size is reduced. With decreasing droplet size, the degree of shear thinning in concentrated emulsions is also enhanced.
It should be noted that the rheological measurements by Rajinder Pal were performed in low-Reynolds-number shear flow~\cite{Pal2000ShearVB}.
The similar shear thinning effect observed for emulsions both in low-Reynolds-number regime and in turbulent state may indicate some similarity for the underlying mechanism, which deserves more investigation.

The above discussion of the effective viscosity of an emulsion reveals some basic rheological behaviors of a turbulent emulsion. However, it remains to be further studied the detailed mechanism by which dispersed phase affects the global transport of systems for high volume-fractions.

\subsection{Phase inversion in dense turbulent emulsion systems}

\begin{figure}
    \centering
    \includegraphics[width = 1\textwidth]{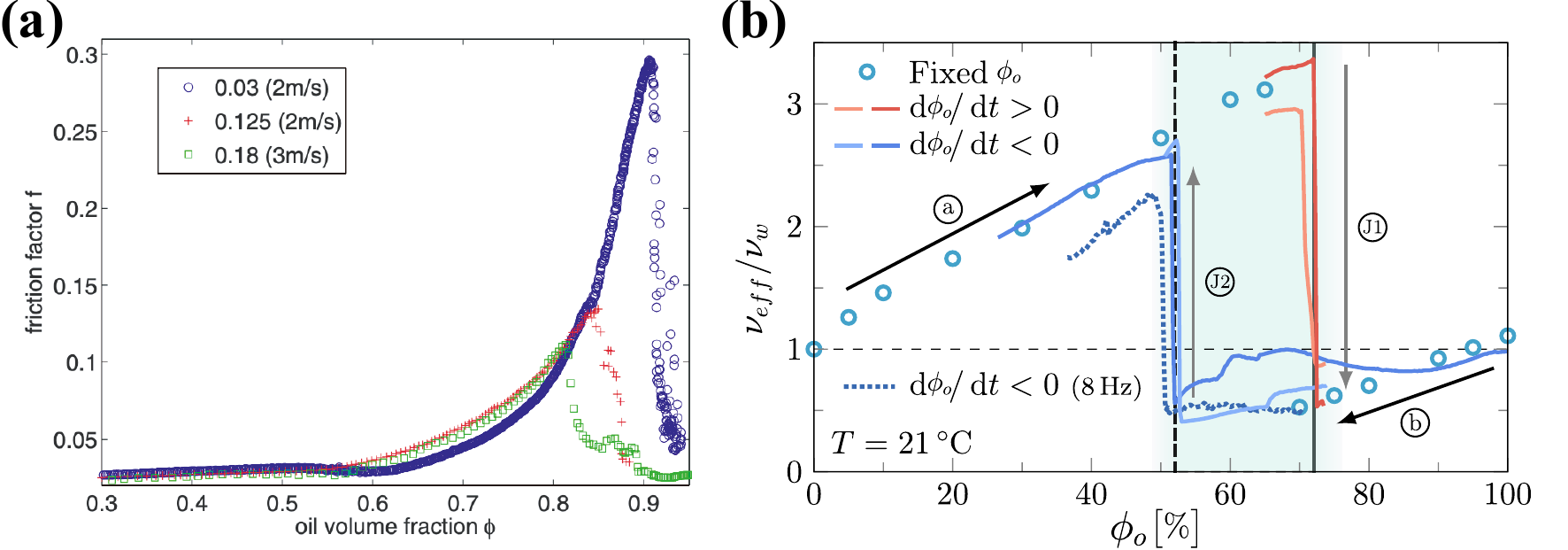}
    \caption{
    \textcolor{black}{
    (a) The evolution of the friction factor with oil volume-fraction in a pipe flow. The water was continuously added into the oil in three experiments at a mixture velocity of 2~m/s and 3~m/s and various injected phase volume fractions $\chi$ (0.03, 0.125, 0.18). Figure adapted from Piela {\it et al.}~\cite{piela2009phenomenological}.
    Note that the $\phi$ is the same as the oil volume-fraction $\phi_{o}$ defined in this review.
    (b) The normalized effective viscosity versus the oil volume fraction in Twente Taylor--Couette system.
    The circles denote results with distinct $\phi_{o}$. The solid lines represent the continuous measurements, in which the fraction is changed gradually during the experiments at a fixed rotation frequency ($\omega_{i}/(2\pi) = 17.5$~Hz), while the dotted line shows the experiment performed at a lower frequency of $\omega_{i}/(2\pi) = 8$~Hz.
    The ambivalence region is shown using the shaded band between the dashed vertical lines, bounded by the two-phase inversion events, $J_{1}$: W/O to O/W and $J_{2}$: O/W to W/O. Figure adapted from Bakhuis {\it et al.}~\cite{bakhuis2021catastrophic}.}. 
    }
    \label{fig4} 
\end{figure}

Apart from the properties of an emulsion at constant volume-fractions of either the O/W or W/O type, the dynamic process of the change between these two types (i.e.\ phase inversion) is also interesting.
Phase inversion is the phenomenon by which the dispersed phase in an emulsion spontaneously inverts to become the continuous phase and vice versa, which has important implications in a range of industrial processes, such as liquid-liquid extraction, and drag reduction in oil/water pipe flow~\cite{yeo2000phase}.
There are many pathways to induce the phase inversion, of which changing the water/oil ratio is a common one, and the induced emulsion inversion is usually called catastrophic phase inversion, due to its very short timescale.
Typically, the dynamic evolution of the phase inversion is accompanied by a significant change of the emulsion properties, including its morphology, stability, and rheology~\cite{perazzo2015phase,bakhuis2021catastrophic}.
The subject of catastrophic phase inversions has attracted much interest in the past decades.

The phase inversion of emulsions has been widely investigated in stirred vessels in many studies, where the role of the stirring speed, temperature, liquid properties (for example, density, viscosity, and interfacial tension), geometry, and materials of the impeller and of container are discussed~\cite{zambrano2003emulsion,mira2003emulsion,tyrode2005emulsion,rondon2007emulsion,perazzo2015phase}.
It is found that if the density difference between the two liquids is high, a phase inversion is promoted because of the increasing relative velocities between dispersed and continuous phases in an agitated vessel~\cite{yeo2000phase}.
Piela et al.~\cite{piela2006experimental,piela2008phase,piela2009phase,piela2009phenomenological} have performed a series of experiments to investigate the phase inversion in a pipe flow, focusing on the inversion mechanism and phase inversion critical volume-fraction (the dispersed phase volume fraction  where inversion occurs).
During the continuous phase inversion experiments, they began with a flow of continuous phase and gradually injected the dispersed phase into the flow until the phase inversion occurred, while fixing the flow rate of mixture in the pipe. 
They analyzed the images of the morphological transitions of the dispersed phase during the phase inversion, which provides significant details for understanding the phase inversion mechanism~\cite{piela2008phase}.
Moreover, a strong increase of the friction factor $f$ (pressure drop) was measured during the inversion (see figure~\ref{fig4}(a)).
Their results showed that the friction factor does not depend on the injected phase volume-fraction $\chi$ (ratio of injected rate of the dispersed phase to the flow rate of mixture) in the pipe before the phase inversion, while the critical volume-fraction corresponding to the phase inversion depends on the injected phase volume-fraction~\cite{piela2006experimental}.
However, it was found that the critical volume fraction is not dependent on Reynolds number, Froude number, Weber number and also on the dispersed phase injection velocity if the mixture velocity is sufficiently large, which also supports the validity of the Ginzburg--Landau mean-field theory for phase inversions, exploited by Piela et al.~\cite{piela2009phenomenological}.

Although the above-mentioned studies provide many insights on phase inversion, much research is still urgently required to understand the phase inversion process and its underlying mechanisms.
Motivated by this, a recent study has been performed to explore the catastrophic phase inversion of the emulsion in a high Reynolds number TC flow~\cite{bakhuis2021catastrophic}.

First, each experiment was performed at a constant oil fraction $\phi_{o}$ and the two liquids were well mixed in a turbulent flow. 
As shown with blue circles in figure~\ref{fig4}b, the effective viscosity of the emulsion is found to increase with increasing oil volume-fraction until a dramatic drop at an oil fraction above $65\%$, which corresponds to a catastrophic phase inversion. Note that the decreasing trend of the effective viscosity with decreasing the oil volume-fraction in the right branch ($\phi_{o} \geq 70\%$) could be due to that the relatively large density difference between oil and water used in these experiments affects the distribution of the two liquids in the system. 
For the purpose of detailed investigation of the phase inversion, the dispersed oil (or water) phase was gradually injected into the system to quasi-statically increase the oil (or water) fraction, during which the instantaneous value of oil volume-fraction was calculated based on the assumption of injection of liquid into a homogeneously mixed emulsion.
The catastrophic phase inversion was observed for both the cases of $d\phi_{o}/dt > 0$ (red lines, at $\phi_{o} = 72\%$) and the cases of $d\phi_{o}/dt < 0$ (blue lines, at $\phi_{o} = 52\%$), where the drag of the system experiences a sharp jump (around $35\%$ in torque).
Therefore, an ambivalence region $52\% \leq \phi_{o} \leq 72\%$ is found in the emulsion system (see the shaded blue band in figure~\ref{fig4}b), which means both O/W and W/O states can exist in this fraction range, and the state of the flow is determined by its current volume-fraction and its history. 
This opens possibilities for drag reduction for the transport of emulsions by actively controlling the emulsion state, which has applications in oil recovery and the petrochemical industry.
In addition, it has been found that in turbulent pipe flows the width of the ambivalence region only depends on the ratio of the injection rate of the dispersed phase to the total flow rate~\cite{piela2006experimental,piela2008phase}. 
However, in this system, as the turbulent strength (i.e. Taylor number or equivalently the Reynolds number) is decreased at a given injection rate of the dispersed phase ($12.5$ mL/s), the width of the ambivalence region slightly increases (shown from dashed to dotted boundaries in figure~\ref{fig4}b).
This suggests a delay of the phase inversion for both O/W and W/O, which is inconsistent with predictions based on an extended Ginzburg--Landau model that the ambivalence region width is independent of the Taylor number~\cite{piela2009phenomenological}.

The results above show that the catastrophic phase inversion of an emulsion can be determined by the dramatic change of the drag (effective viscosity) of the system. However, the detailed dynamics of both the local droplet structures and global drag properties during the ``sudden" inversion are far from being fully understood and deserve further investigation, experimentally, numerically, and theoretically.

\section{Summary and outlook}

Emulsions are common both in nature and in various industrial processes, including enhanced oil recovery, chemical engineering, and food processing. However, the understanding of the physical mechanism of the emulsions, particularly turbulent emulsions, is still limited.
In this paper, we review some recent studies on turbulent emulsions in TC flow, mainly focusing on both the statistics and breakup mechanism of dispersed droplets, and the transport properties (effective viscosity).

First, at a low oil volume-fraction of turbulent emulsions, the experimental results show that the PDF of the droplet sizes follows a log-normal distribution for various $\text{Re}$, suggesting that the droplet generation could be determined by a cascade process of random fragmentation. 
It is found that the normalized droplet size has a $-1.18$ scaling dependence on the Reynolds number. This scaling behavior can be derived using either KH theory with the energy dissipation rate or Levich theory. However, based on the analysis of the droplet Weber number, it's found that the energy dissipation rate in the bulk flow is not enough to induce the breakup of the droplets of typical size in the system. It is concluded that droplet fragmentation, which gives the typical droplet size, occurs within the boundary layer and is determined by the dynamic pressure across the droplet size due to the velocity gradient of the mean flow, as proposed by Levich.

For high volume-fractions turbulent emulsions, the coupling interaction between the local droplet structures and global transport (rheology) characterizes the dynamics of the emulsion system. 
The effective viscosities of an emulsion for various oil volume-fractions are calculated based on a method based on the recent developments in turbulent TC flow. It is found that the effective viscosity increases with increasing oil volume fraction, which is consistent with the numerical results obtained in planar Couette flow and the overall trend is qualitatively similar to the rheological behavior of suspensions of solid particles. 
However, the dispersed droplets in turbulent emulsions are expected to show richer dynamics when compared to the suspension of solid particles.
Furthermore, the effective viscosity decreases with increasing $\text{Re}$ (or, the shear stress), indicating a shear thinning effect. This shear thinning, which becomes stronger at higher-volume fractions, can be quantitatively described by the classical Herschel--Bulkley model through the dependence of the flow index on the volume-fraction. 
Similar shear thinning effects have also been reported for both the emulsions in a low-Reynolds-number flow and the suspensions of deformable particles under a steady shear flow.

Finally, we review recent experimental results on phase inversions of emulsions in a stirred vessel, pipe flow, and high Reynolds number TC flow. As the dispersed phase volume fraction is increased to a critical value, a catastrophic phase inversion occurs, accompanied by a sudden and dramatic change of the rheological properties (effective viscosity) of the emulsion. An ambivalence region of inversion is observed in the system, opening possibilities for drag reduction for the transport of emulsions by actively controlling the emulsion state.
Moreover, the width of the ambivalence region is found to slightly increase for decreasing Taylor (Reynolds) number of the flow, which is in contrast to the prediction of extended Ginzburg--Landau theory.

Although the results summarized above have provided many insights into the emulsion in a TC flow, more studies are needed for a comprehensive understanding of the dynamics in the system.
Among them is the physical mechanism of the interaction between local droplet structures and the global transport in turbulent emulsion, particularly at high volume-fractions.
In addition, more experimental, theoretical, and numerical investigation is needed on the detailed dynamics of a catastrophic phase inversion, including the instantaneous morphological changes of the dispersed phase and the drag properties.

\vskip6pt

\enlargethispage{20pt}


\dataccess{This article has no additional data.} 

\aucontribute{LY, CW, SH, and CS designed the paper structure; LY drafted the paper; CW, SH, and CS revised the paper.}

\competing{The authors declare no competing financial or non-financial interests. .}

\funding{This work was supported by the Natural Science Foundation of China under grant no. 11988102 and 91852202, the Tencent Foundation through the XPLORER PRIZE, and an European Research Council Starting Grant (SH) and the Netherlands Organisation for Scientific Research through the Multiscale Catalytic Energy Conversion research center (SH).}

\ack{We thank D. Lohse, F. Risso, and F. Toschi for various discussions over the years.}


\end{document}